\documentclass[aps,prd,twocolumn, 10pt,longbibliography]{revtex4-1}

\usepackage{graphicx}% Include figure files
\usepackage{dcolumn}% Align table columns on decimal point
\usepackage{bm}% bold math
\usepackage{siunitx}
\usepackage[colorlinks=true,citecolor=blue,linkcolor=blue]{hyperref}
\usepackage{enumitem}
\setlist{nosep}

\hypersetup{
pdfauthor = {Brendan J. Shields},
colorlinks = true, linkcolor = blue, urlcolor=blue, bookmarksnumbered =  true}

\newcommand{\ket}[1]{\left|#1\right>}

\newcommand{\iwg}{I_{\mathrm{wg}}} % Waveguided power
\newcommand{\ibd}{I_{\mathrm{bd}}} % Power radiated by an NV in uniform bulk diamond.
\newcommand{\ina}{I_{\mathrm{na}}} % Objective-coupled power.

\begin{document}

\title{Parabolic diamond scanning probes for single spin magnetic field imaging}% Force line breaks with \\
\author{Natascha Hedrich}
\affiliation{Department of Physics, University of Basel, Klingelbergstrasse 82, Basel CH-4056, Switzerland}
\author{Dominik Rohner}
\affiliation{Department of Physics, University of Basel, Klingelbergstrasse 82, Basel CH-4056, Switzerland}
\author{Marietta Batzer}
\affiliation{Department of Physics, University of Basel, Klingelbergstrasse 82, Basel CH-4056, Switzerland}
\author{Patrick Maletinsky}
\affiliation{Department of Physics, University of Basel, Klingelbergstrasse 82, Basel CH-4056, Switzerland}
\author{Brendan J. Shields}
\email[]{brendan.shields@unibas.ch}
\affiliation{Department of Physics, University of Basel, Klingelbergstrasse 82, Basel CH-4056, Switzerland}

\date{\today}% It is always \today, today,

\begin{abstract}
Enhancing the measurement signal from solid state quantum sensors such as the nitrogen-vacancy (NV) center in diamond is an important problem for sensing and imaging of condensed matter systems. Here we engineer diamond scanning probes with a truncated parabolic profile that optimizes the photonic signal from single embedded NV centers, forming a high-sensitivity probe for nanoscale magnetic field imaging. We develop a scalable fabrication procedure based on dry etching with a flowable oxide mask to reliably produce a controlled tip curvature. The resulting parabolic tip shape yields a median saturation count rate of \SI[separate-uncertainty]{2.1\pm0.2}{MHz}, the highest reported for single NVs in scanning probes to date. Furthermore, the structures operate across the full NV photoluminescence spectrum, emitting into a numerical aperture of 0.46 and the end-facet of the truncated tip, located near the focus of the parabola, allows for small NV-sample spacings and nanoscale imaging. We demonstrate the excellent properties of these diamond scanning probes by imaging ferromagnetic stripes with a spatial resolution better than \SI{50}{\nm}. Our results mark a 5-fold improvement in measurement signal over the state-of-the art in scanning-probe based NV sensors.
\end{abstract}

\maketitle

\section{Introduction}

Understanding condensed matter systems at the nanoscale is increasingly important for a wide range of topics ranging from nanomagnetism to structural imaging in biology.  Sensing and imaging the tiny sample volumes in such systems requires high sensitivity and high spatial resolution.  Sensors based on solid state defects such as the nitrogen-vacancy (NV) center in diamond have emerged as ideal platforms for addressing such nanoscale phenomena~\cite{casola_CondMatNVReview_2018}, owing to the potential for sensor volumes approaching atomic dimensions.  The NV contains a single electronic spin associated to an atomic-scale lattice defect, which can be initialized and read out optically~\cite{gruber_ODMR_1997,doherty_NVreview_2013}, and the diamond host material can be readily integrated into scanning probe devices~\cite{maletinsky_robustsensor_2012,appel_fabrication_2016}.  

These unique properties have enabled the quantitative imaging of nanoscale systems such as skyrmions~\cite{jenkins_single-spin_2019,dovzhenko_magnetostatic_2018,gross_skyrmion_2018}, domains in antiferromagnets~\cite{gross_real-space_2017,appel_nanomagnetism_2019}, electron transport in graphene~\cite{ku_viscous_2019,jenkins_ohmictransport_2020}, and magnetism in 2D materials~\cite{thiel_probing_2019}.  Detection of weak signals originating from nuclear spins of single proteins \cite{lovchinsky_nuclear_2016} or 2D materials \cite{lovchinsky_nqr_2017} has been demonstrated in a static sensor configuration, but scanning probe imaging of such systems requires higher sensitivity.  The remarkable potential of scanning NV magnetometry and the wide range of even more challenging applications that lie ahead point to the need for improved diamond scanning probe technology, which is the focus of this work.

Two aspects are key to the performance of NV-based nanoscale sensors and will be addressed in this work.  First, an NV center in close proximity to the diamond surface is required to minimize NV-sample separation for optimal spatial resolution, and to maximize the magnetic signal from nanoscale sample volumes. Second, a high flux of detected photons from NV photoluminescence (PL) is required to efficiently distinguish between the bright $m_s = \ket{0}$ and the dark $m_s = \ket{\pm 1}$ NV spin states, for optimal magnetic sensitivity. The high refractive index of the diamond host (n=2.4) presents a challenge for the second requirement, but also offers a natural route to engineer photonic structures that maximize PL detection. Many approaches have been taken to optimizing collection efficiency via photonic engineering of diamond, including solid immersion lenses~\cite{hadden_firstDiamondMicroSIL_2010,jamali_111diamondSIL_2014}, diffraction optics~\cite{li_efficient_2015,huang_metalens_2019}, dielectric antennas~\cite{riedel_dielectricantenna_2014}, parabolic reflectors~\cite{wan_efficient_2018}, and waveguiding structures~\cite{hausmann_pillarfabrication_2010,momenzadeh_nanoengineered_2015,mouradian_scalable_2015}. Of these, the parabolic reflector geometry has yielded the highest flux of photons, and is a promising candidate for improving the sensitivity of NV magnetometry experiments. However, it remains an outstanding challenge to employ this highly efficient photonic structure in a scanning probe configuration for nanoscale magnetometry, in part because of the large standoff distance between the diamond surface and the NV position at the focal point of the paraboloid.

In this work, we adapt the parabolic reflector to a scanning probe geometry by truncating the paraboloid apex (Fig.~\ref{fig:ConceptAndSimulationFigure}a), which allows for small NV-sample spacing while maintaining the paraboloid's advantageous photonic properties. This yields a near-surface NV in a high collection efficiency, broadband, waveguided device, thereby satisfying both of the requirements outlined above.

\section{Concept and simulations}

\begin{figure}[b]
\includegraphics[width=8.6cm]{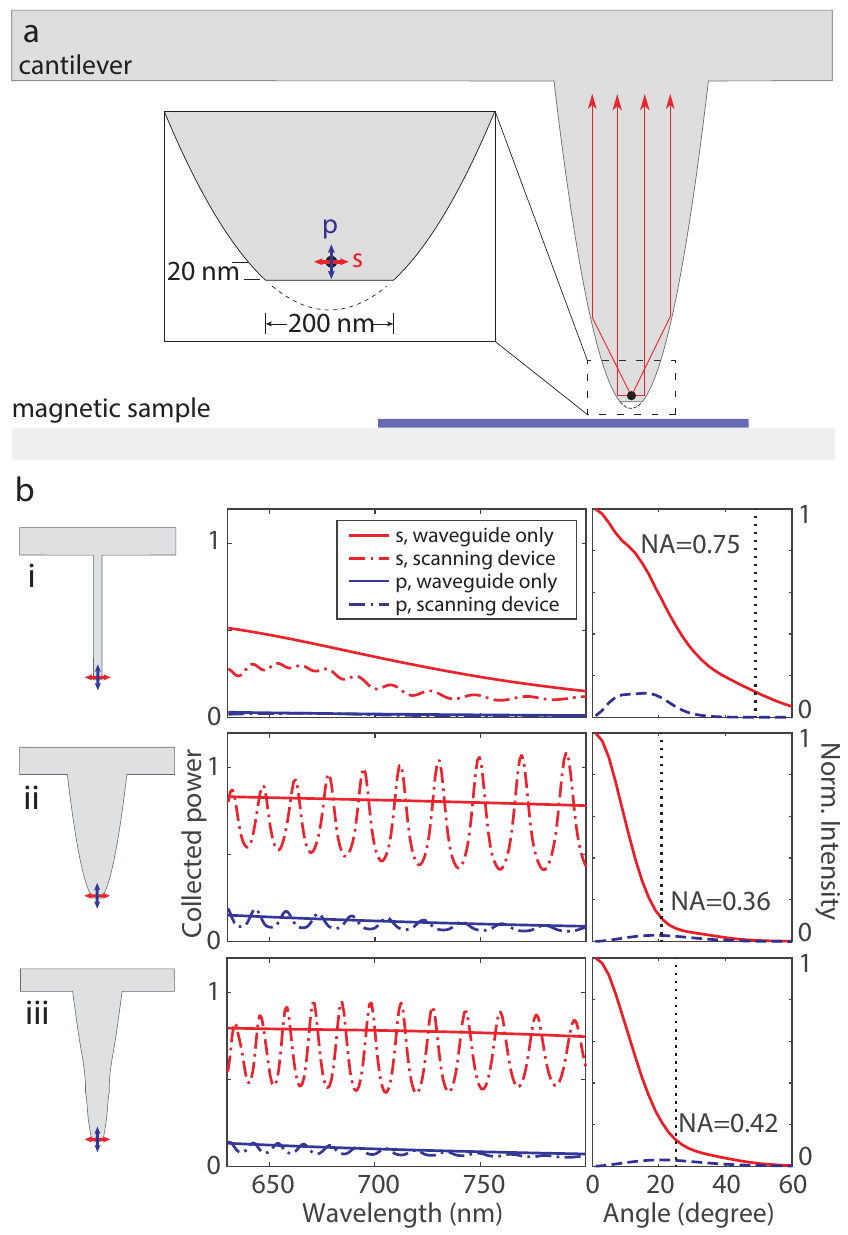}
\caption{\label{fig:ConceptAndSimulationFigure} \textbf{(a)} Diamond parabolic scanning probe concept showing waveguiding of the emission of an NV close to a magnetic surface. The inset shows the length scales involved in the parabolic design.  \textbf{(b)} Simulated device performance for (i) cylindrical, (ii) parabolic and (iii) fabricated nanopillars.  The left panel shows the ratio of waveguided power emitted into $\mathrm{NA} = 0.8$, to the power emitted by a bulk dipole. This ratio is shown for the pillar (waveguide-wg) alone (solid lines), and the entire simulated device (dashed lines) as a function of wavelength for $s$-polarized (red) and $p$-polarized (blue) dipoles.  All powers are normalized to the total power emitted by a dipole embedded in bulk diamond material. The right panel in each case shows the normalized, integrated NV photoluminescence emitted from the device as a function of the polar exit angle for the $s$-polarized (red) and $p$-polarized (blue) dipoles.}
\end{figure}

Our approach adapts the well-established concept of a diamond parabolic reflector~\cite{wan_efficient_2018} to a pillar-on-cantilever geometry~\cite{maletinsky_robustsensor_2012} that we incorporate into an atomic force microscope probe~\cite{appel_fabrication_2016} for scanning magnetic field imaging (Fig.~\ref{fig:ConceptAndSimulationFigure}(a)).  Conventional scanning probes employ a cylindrical~\cite{maletinsky_robustsensor_2012} or tapered pillar~\cite{zhou_scanning_2017} geometry that acts to waveguide the NV PL.  Our improved design replaces this by a truncated diamond paraboloid having an NV at its focal point.  The intuition guiding our choice of a parabolic geometry can be seen through a ray optics picture (Fig.~\ref{fig:ConceptAndSimulationFigure}(a)), where total internal reflection at the parabolic surface collimates the emission into a unidirectional output mode, resulting in improved waveguiding of the NV emission.  The truncated design with flat end facet minimizes the distance of the paraboloid's focal point, and hence the NV, from the sample. 

To fully understand the performance characteristics of our design, we go beyond the ray optics picture by simulating the device with a finite-difference time-domain solver (Lumerical), taking a cylindrical pillar waveguide as a basis of comparison~\cite{hausmann_pillarfabrication_2010}.  Both cylindrical and parabolic devices are set to have a facet diameter of \SI{200}{\nm}, which approximately corresponds to the minimal diameter that still supports an optical mode with strong confinement to the diamond. Furthermore, we examine dipole sources oriented perpendicular ($s$-polarized) and parallel ($p$-polarized) to our pillar axis. The allowed electric dipole transitions for the NV correspond to dipoles lying in the plane perpendicular to the NV symmetry axis, which is oriented along one of the [111] crystal directions.  For a given crystal orientation there are then four possible NV directions. 

We assess the device performance on the basis of two key metrics: outcoupled power $\ina$ within the 0.8 numerical aperture (NA) cone of a microscope objective, and the directionality of the NV emission.  All powers are normalized to the power radiated by a dipole in uniform bulk diamond, $\ibd$.  Note that $\ina/\ibd$ is related to the collection efficiency $\eta$ of the device, but also includes the near-field effect of the diamond surface and the Purcell effect due to reflection off the back side of the cantilever, which modify the radiative decay rate of the dipole. For $p$-polarized dipoles or for cylindrical devices the difference is quite large, however, for an $s$-polarized dipole in a parabolic device, $\ina/\ibd$ and $\eta$ differ by only 1\% on average over the NV emission band.   

We first simulate the complete device, consisting of pillar and cantilever (dashed lines in Fig.~\ref{fig:ConceptAndSimulationFigure}(b)), for which an $s$-polarized dipole (red lines) in the cylindrical device leads to a value of  $\ina^{\rm{cyl}}/\ibd = 0.18$ (averaged across the \SI{630}{\nm} to \SI{800}{\nm} NV emission band), while the same dipole in the parabolic device gives $\ina^{\rm{para}}/\ibd = 0.68$, a nearly factor of 4 improvement.  The interference fringes in the spectrum result from reflections of NV emission at the diamond-air interface at the back of the cantilever.  To isolate the contribution of the parabola from this interference signal, we perform a second simulation in which the waveguide section is terminated into the perfectly absorbing wall of the simulation space to limit reflections.  The normalized power into the waveguide $\iwg/\ibd$ for this case is shown by the solid lines in Fig.~\ref{fig:ConceptAndSimulationFigure}(b), and exhibits broadband performance with an average value of 0.81 over the NV emission band, which represents the upper limit for $\ina/\ibd$ that could be achieved through antireflection coating of the cantilever backside.  We note that the simulated performance of a fabricated device, shown in Fig.~\ref{fig:ConceptAndSimulationFigure}(b-iii), closely matches that of the ideal parabolic pillar, with $\ina^{\rm{dev}}/\ibd = 0.64$. 

We additionally consider the far-field emission pattern, which we extract from the full structure simulation.  The emission intensity from the device is plotted as a function of polar angle (Fig.~\ref{fig:ConceptAndSimulationFigure}(b) right column), and shows that the larger aperture of the parabolic device concentrates the far-field emission within a narrow NA of 0.36.  The cylindrical pillar, on the other hand, undergoes significant diffraction due to its wavelength-scale aperture, resulting in a much larger NA of 0.75.

We note that in the case of a $p$-polarized dipole (Fig.~\ref{fig:ConceptAndSimulationFigure}(b), blue lines), the outcoupled power is in all cases strongly suppressed due to the near-field diamond-air interface~\cite{lukosz_light_1977} and poor overlap of the NV emission with the waveguide mode.  It is clear from these findings that an $s$-polarized dipole is optimal, which would be the case if the NV axis were aligned to the pillar axis~\cite{davies_NVoptical_1976,rohner_111-oriented_2019}. 

\section{Fabrication}

\begin{figure}[b]
\includegraphics[width=8.6cm]{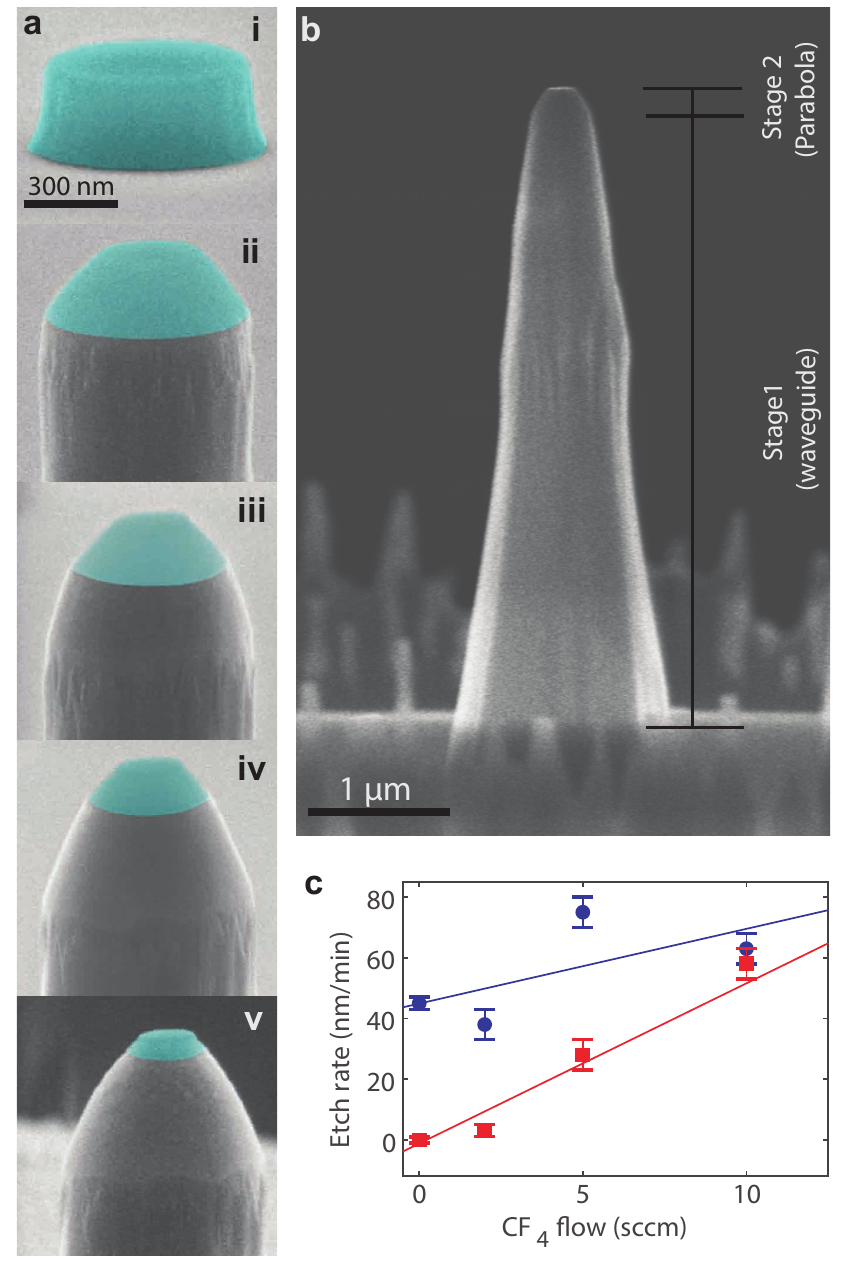}
\caption{\label{fig:Fabrication}  (a) SEM images of fabrication sequence.  (i) 300-nm thick FOX-16 disk.  (ii) After the first stage of O$_2$ etching, the mask is eroded at the edges, leaving a trapezoidal cross section.  (iii-v) Subsequent etching with increasing CF$_4$ flow results in a controlled mask erosion which leads to a parabolic diamond surface.  (b) SEM image of completed pillar, consisting of parabolic tip and tapered waveguide.  (c) Etch rates of diamond (blue circles) and FOX mask (red squares) vs.\,CF$_4$ concentration, with linear fits (solid lines). }
\end{figure}

To fabricate a pillar with parabolic curvature, we developed a procedure based on dry-etching with a flowable oxide mask in which we vary the mask erosion rate relative to the diamond etch rate, yielding precise control over the curvature of the final diamond device.  We begin fabrication with a high-purity type-IIa diamond prepared with a layer of implanted NV centers and pre-patterned with an array of cantilevers etched to a depth of \SI{2}{\micro\meter}~\cite{SI}. As an etch mask for the parabolic pillars, we then pattern 1-$\mu$m diameter discs in a $\sim$300-nm thick layer of flowable oxide (FOX-16, Dow Corning) via electron beam lithography onto the cantilevers (Fig.~\ref{fig:Fabrication}(a)).  Etching of the pillars then consists of two ICP-RIE (Sentech) stages. 
\begin{table} [h]
	\begin{tabular}{cccccccc}
		\hline 
		Stage & O$_2$ Flow & CF$_4$ Flow & Pressure & ICP & Bias & Time\\
		& (sccm) & (sccm)& (Pa) &(W)&(V)&(s)\\ \hline\hline
		& 50 &0& 0.5 & 500 & 110 & 240 \\
		1&&&&&&\\
		&50 & 2& 0.5 & 500 & 40 & 4\\\hline\\
		2 & 50 & 2-10 & 0.5 & 500 & 40 & variable \\ \hline\hline
	\end{tabular}
	\caption{\label{tab:Etching}Summary of the plasma parameters used for etching the parabolic diamond tip.}
\end{table} 

The first stage etches a tapered diamond waveguide, and consists of \SI{240}{s} steps of O$_2$ etch chemistry alternating with \SI{4}{s} steps of O$_2$ and CF$_4$ to clean off resputtered material from the walls of the device~\cite{yamada_cycle_2007}. The two etch steps are repeated a total of nine times to achieve a $\sim$\SI{6}{\micro\meter} tapered pillar. At the end of this stage, the mask has a trapezoidal cross section with a base diameter of \SI{900}{\micro\meter} (Fig.~\ref{fig:Fabrication}(a-ii)).  
For the second stage, we introduce CF$_4$ throughout the etch at increasing flow rates for successive steps. This procedure controllably erodes the FOX mask in proportion to CF$_4$ concentration.  Together with the trapezoidal cross section, this allows us to tune the angle of the walls (Fig.~\ref{fig:Fabrication}(a-iii--v)) by controlling the relative etch rate of FOX and diamond (Fig.~\ref{fig:Fabrication}(c)). The details of these stages are outlined in Tab.~\ref{tab:Etching}. A typical final device is shown in Fig.~\ref{fig:Fabrication}(b) and exhibits a parabolic tip section with a $\sim200$-nm flat end facet and a long tapered waveguide section.  Simulations of the final device profile indicate similar performance characteristics to the initially targeted parabolic geometry (Fig.~\ref{fig:ConceptAndSimulationFigure}(b-iii)).  Following the parabolic etch, a deep etch from the back side of the diamond releases the cantilevers for assembly into scanning probes following~\cite{appel_fabrication_2016}.

\section{Device characterization}

\begin{figure}[b]
	\includegraphics[width=8.6cm]{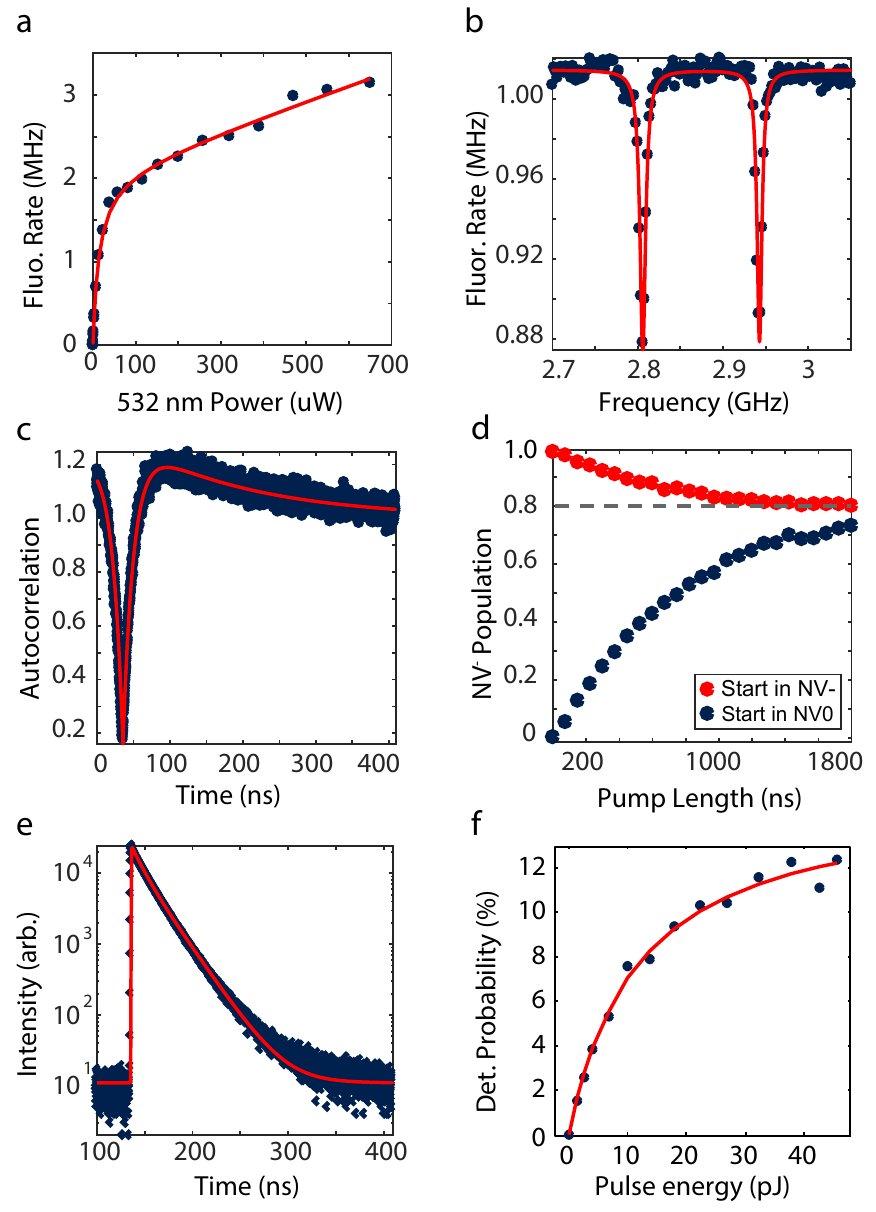}
	\caption{\label{fig:Char} Representative measurements to characterize the optical and NV-related properties of the fabricated devices. All measurements, save \textbf{f} are taken from the same device. \textbf{(a)}  Saturation curve taken with \SI{532}{\nm} excitation resulting in $I_{\mathrm{PL}}$=\SI[separate-uncertainty]{2.1\pm0.2}{\MHz} and $P_{\mathrm{sat}}$=\SI[separate-uncertainty]{14\pm3}{\micro W} . \textbf{(b)} ODMR of a single NV in the scanning probe taken at saturation power with the dips fit by a Lorentzian (red).  \textbf{(c)} Second order correlation function (g$^{(2)}(\tau)$) of the NV, revealing strong antibunching (g$^{(2)}(0)=0.16$), hence the presence of a single NV. \textbf{(d)} NV$^-$ population measured by \SI{590}{\nm} excitation after starting in NV$^0$ (blue points) or NV$^-$ (red points) for different \SI{532}{\nm} pulse lengths, converging to a steady-state NV$^-$ population of $0.80\pm0.2$.	\textbf{(e)} Radiative lifetime of the NV excited state fitted by an exponentially modified Gaussian function (red line) resulting in a lifetime of \SI[separate-uncertainty]{21\pm1}{\ns} for the m$_s = \ket{0}$ state.\textbf{(f)} Overall detection efficiency of a second fabricated structure as a function of the \SI{532}{\nm} excitation power with a saturation fit (in red).}
\end{figure}

We characterize our devices in a homebuilt confocal microscope, equipped with CW excitation lasers at \SI{532}{\nm} and \SI{594}{\nm}, a tunable supercontinuum picosecond pulsed laser (SuperK Extreme, NKT Photonics), and a gold wire loop mounted on a translation stage to deliver microwaves for spin manipulation to the sample.  A bar magnet supplies an external magnetic field for Zeeman splitting of the $m_s=\ket{\pm1}$ electron spin states.  A single photon counting module (SPCM) (AQRH-33, Excelitas) detects PL, with a 45/55 beamsplitter and second SPCM being inserted for auto-correlation measurements.  We characterize the angular distribution of emission with a back focal plane imaging system, as described in~\cite{rohner_masterthesis_2013}.

\subsection{Single NV characterization}

\begin{figure}[h]
	\includegraphics[width=8.6cm]{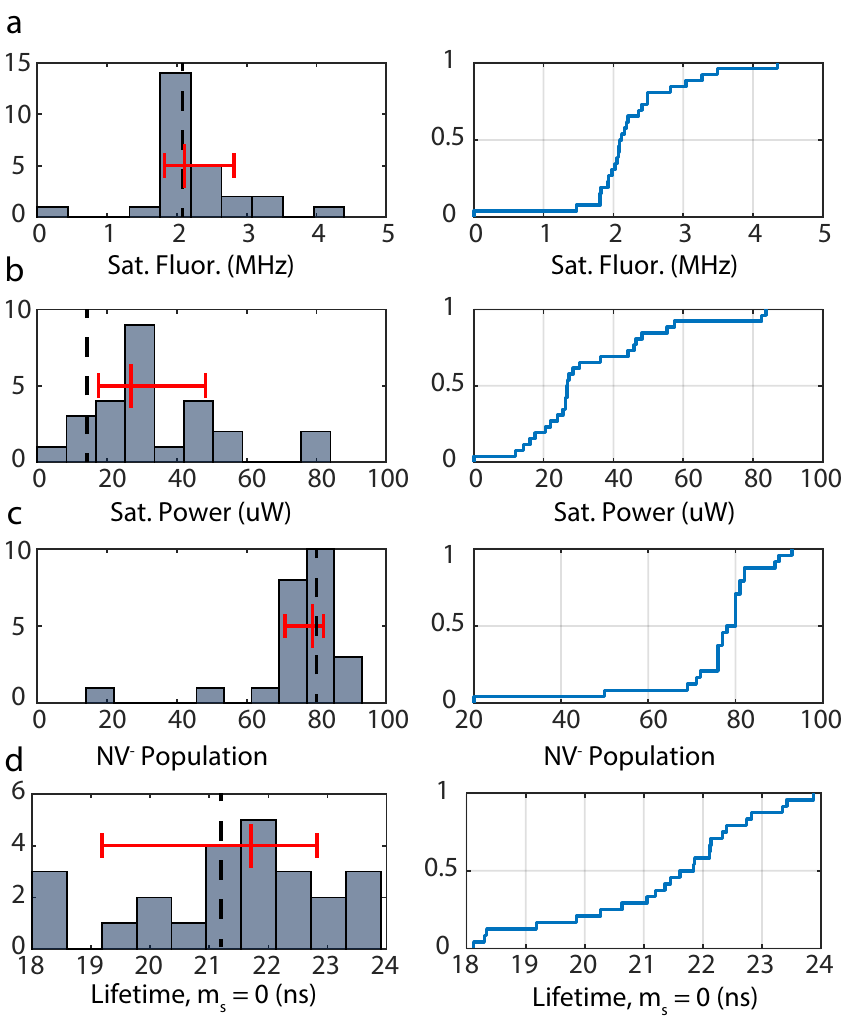}
	\caption{\label{fig:Stats} Combined statics of all 25 investigated single-NV structures depicted histogramically (left column) and with the Cumulative Distribution Function (CDF) (right column). The 1$\sigma$ band is given by the red horizontal line and the median is depicted with a vertical red line. The values for the device investigated in Fig.~\ref{fig:Char} are marked by the vertical, dashed, black lines. \textbf{(a)} Saturation PL, \textbf{(b)} Saturation Power, \textbf{(c)} Steady-state NV$^-$ population, \textbf{(d)} Lifetime of the NV $m_s = \ket{0}$ state.}
\end{figure}

We begin by searching for devices containing a single NV via optically detected magnetic resonance (ODMR) spectra~\cite{gruber_ODMR_1997}.  Here, the bar magnet is positioned to generate a unique Zeeman splitting for each of the 4 NV orientations, and the microwave frequency is swept while recording the PL from the NV. We identify pairs of resonance dips in the PL, resulting from the transitions between the $\ket{0}\rightarrow\ket{\pm1}$ states, which then indicate the number of NVs in the structure. From a writefield of 288 devices we identify 108 as containing a single pair of ODMR lines, 60 as containing multiple ODMR pairs, and the remainder exhibiting only background fluorescence, corresponding to an average of \SI{0.91\pm0.05} NVs per device.  Of the 108 single-ODMR pair devices, we selected 36 from across the writefield for further characterization. For each device, we assess its performance by recording a PL saturation curve, PL autocorrelation ($g^{(2)}(\tau)$), photoionization rates at saturation, and a PL lifetime measurement, as described below. The complete dataset for a representative device is shown in Fig.~\ref{fig:Char}(a--e).  We use the ODMR, autocorrelation, and photoionization measurements (Fig.~\ref{fig:Char}(b-d)) to confirm that 25 of the 36 selected devices contain a single NV center.  The distributions for each measurement over the full set of 25 single-NV devices are shown in Fig.~\ref{fig:Stats}.

The PL saturation measurement (Fig.~\ref{fig:Char}(a)) is our primary figure of merit for brightness, and consists of the steady-state PL rate $I$ as a function of CW \SI{532}{\nm} laser power $P$, which we fit to $I(P)=\frac{I_{\mathrm{PL}}P}{P+P_{\mathrm{sat}}} + I_{\mathrm{bkgd}}P$ to extract the saturated PL rate $I_{\mathrm{PL}}$, saturation power $P_{\mathrm{sat}}$, and background $I_{\mathrm{bkgd}}$.  The distributions of $I_{\mathrm{PL}}$ and $P_{\mathrm{sat}}$ are shown in Fig.~\ref{fig:Stats}(a,b) for all 25 single-NV devices, with the median  $I_{\mathrm{PL}}=$\SI{2.1}{\MHz} and $P_{\mathrm{sat}}=$\SI{27}{\micro W}. These are, to our knowledge, the highest published count rates for NV centers in a scanning probe geometry. Furthermore, the low saturation powers we find are generally advantageous, especially in cryogenic conditions, where high laser powers may lead to heating. 

Steady-state PL is the primary figure of merit for brightness, but it averages over fluctuations due to photoionization of the bright negative charge state (NV$^-$) to the dark neutral charge state (NV$^0$)~\cite{aslam_photoinduced_2013}.  This effect complicates the measurement of photon collection efficiency $\eta$, which would ideally consist of deterministically driving to the NV$^-$ excited state and counting the fraction of collected photons.  We quantify the effect of photoionization by measuring the average NV$^-$ population at $P_{\mathrm{sat}}$ for each device~\cite{bluvstein_identifying_2019,SI} (Fig.~\ref{fig:Char}(d)). The distribution of NV$^-$ populations is shown in Fig.~\ref{fig:Stats}(c) and is clustered around a median of 0.79.  

To isolate $\eta$ from photoionization effects, we therefore perform a saturated, pulsed excitation measurement while controlling for charge state~\cite{SI}.  This consists of an initial charge state measurement, an excitation pulse, and a final charge state measurement.  We count the average detected photons when initial and final charge state are NV$^-$, and subtract the average for initial and final both being NV$^0$, which directly accounts for background photons.  In doing so, we obtain an overall detection efficiency of $\eta=0.12$ for NV$^-$ photons (Fig.~\ref{fig:Char}(f)). This overall efficiency takes into account the parabolic reflector device efficiency $\eta_{\mathrm{dev}}$ and the optical path efficiency $\eta_{\mathrm{setup}}$ which we estimate to be 0.21 (including transmission through optical components, fiber coupling, and SPCM efficiency)~\cite{SI}.  Taking these into account we find a device collection efficiency of $\eta_{\mathrm{dev}} = \eta / \eta_{\mathrm{setup}} = 0.57$.  From our simulations we expect the NV emission to be dominated by $s$-polarization, so that we can take $\eta_{\mathrm{dev}}$ and $\iwg/\ibd$ to be equivalent.  The $\eta_{\mathrm{dev}}$ of 0.57 measured here indicates a device performance that nearly approaches the simulated $\ina^{\rm{para}}/\ibd$ of 0.68.

Finally, we examine the suppression of PL emission in our devices relative to bulk NVs.  We measure the PL lifetime with picosecond pulsed laser excitation and fit the time-resolved PL with a sum of two exponentially modified Gaussian functions (to account for the timing jitter of our SPCMs), extracting a PL lifetime of the m$_s = \ket{0}$ excited spin states~\cite{robledo_spin_2011} (Fig.~\ref{fig:Char}(e)). The $\ket{0}$ state decays primarily radiatively, which is the only part of the decay modified by the photonic properties of our devices.  The distribution of  $m_s = \ket{0}$ lifetimes is presented in Fig.~\ref{fig:Stats}(d), with a median of \SI{22}{ns}, as compared to \SI{13}{\ns} for NV centers in bulk diamond~\cite{robledo_spin_2011}.  In our devices the pillar axis is aligned along the diamond [100] direction due to the starting crystal orientation, resulting in NV emission that is $2/3$ $s$-polarized and $1/3$ $p$-polarized~\cite{epstein_anisotropic_2005,alegre_polarizationselective_2007}. Thus, the observed $m_s = \ket{0}$ lifetimes (Fig.~\ref{fig:Stats}(d)) match our expectation from simulations that the $p$-polarized dipole emission is strongly suppressed in our devices. 
 
\subsection{Emission pattern}

\begin{figure}[b]
	\includegraphics[width=8.6cm]{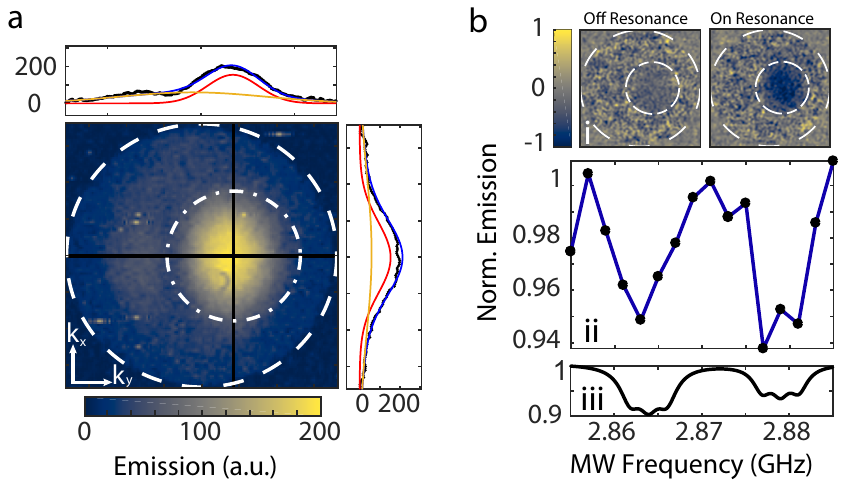}
	\caption{\label{fig:BFP} Back Focal Plane (BFP) imaging of a representative pillar. \textbf{(a)} Waveguided emission into an $\mathrm{NA} = 0.35$ indicated by the dot-dashed line. The dashed line corresponds to an $\mathrm{NA} = 0.8$, given by our objective. Black lines indicate the position of the linecuts at the top and to the right. In these insets, the data is shown in black with the signal portion in red, the background in orange, and the overall fit in blue. \textbf{(b)} BFP imaging while driving the NV spin with near-resonant microwave fields. \textbf{(i)} Differential images off (left) and on (right) resonance, clearly showing the dip in emission when resonant. \textbf{(ii)} Normalized BFP emission integrated over the fitted $\mathrm{NA} = 0.35$. \textbf{(iii)} The ODMR measured in the conventional way, showing the NV resonances at the same frequency as in (ii).}
\end{figure}

In addition to increased collection efficiency, the parabolic design channels NV emission into a narrowly directed output NA.  We confirm this improved angular emission distribution via back focal plane (BFP) imaging of our scanning probes using an apparatus described in~\cite{rohner_masterthesis_2013}.  We image the BFP of our objective onto a CCD, and record an image of the NV PL while exciting the NVs with CW \SI{532}{nm} laser light. Fig.~\ref{fig:BFP}(a) shows the result where the white dashed line indicates the $\mathrm{NA} = 0.8$ limit of the objective aperture.  We fit the NV emission with a 2-dimensional Gaussian distribution plus background~\cite{SI}, and obtain the 1/e$^2$ point (indicated by a white dash-dotted line). From this fit, we extract a best-fit emission NA of $\mathrm{NA_\mathrm{bf}} = 0.35$ for the structure shown in Fig.~\ref{fig:BFP}. Such measurements are performed on a set of 11 structures, with a median of $\mathrm{NA_\mathrm{bf}} = 0.44$, and a 1$\sigma$ confidence interval of $\mathrm{NA_\mathrm{bf}}\in[0.38,0.48]$ . 

The emission from our devices is slightly off-axis due to lateral displacements of the NV in the pillar as well as a slight tilt of the pillar from the sample normal.  From a practical viewpoint, we therefore define an effective numerical aperture $\mathrm{NA}_{\mathrm{eff}}$ referenced to the center of the objective axis, such that $1- 1/e^2$ of the emission is contained within the $\mathrm{NA}_{\mathrm{eff}}$ cone.  We find $\mathrm{NA}_{\mathrm{eff}}=0.46$, again with a 1$\sigma$ confidence interval of $\mathrm{NA}\in[0.40, 0.52]$ in good agreement with $\mathrm{NA_\mathrm{bf}}$. Though not realized in this study, improving the angular distribution of the emission allows for improvements to the collection path, including anti-reflection coating of the diamond cantilever and collection lenses with lower NA, larger working distance, and higher transmission, which would particularly facilitate cryogenic usage of scanning NV magnetometers.

Furthermore, we verify that the BFP emission indeed originates from the NV by performing ODMR while monitoring the BFP signal, as shown in Fig.~\ref{fig:BFP}(b).  The top panel of Fig.~\ref{fig:BFP} shows differential images taken by subtracting the BFP signal without microwave drive from the signal with microwave drive off-resonance (left) and on-resonance (right) to the NV spin transition.  The middle panel shows ODMR measured by integrating the differential image within $\mathrm{NA_\mathrm{bf}}$, and shows very good agreement with ODMR measured with the SPCM (bottom panel).  Comparison of the differential BFP images on- and off-resonance clearly shows that the ODMR contrast is concentrated in the region of the best-fit NA.

\subsection{Nanoscale magnetic field imaging}

\begin{figure}[h]
	\includegraphics[width=8.6cm]{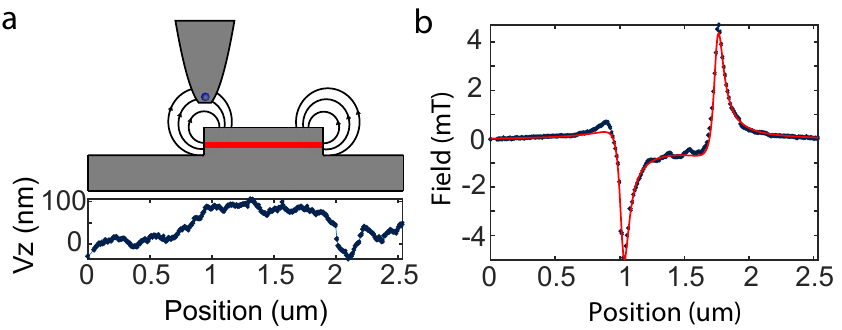}
	\caption{\label{fig:Linecut} \textbf{(a)} Schematic showing the parabolic scanning probe orientation relative to the encapsulated \SI{1}{\nm} thick CoFeB stripe, with the corresponding topography scan.  \textbf{(b)} The field measured through CW ODMR magnetometry as a function of position. The fit (in red) is used to extract the NV-to-sample distance (\SI[separate-uncertainty]{45\pm5}{\nm}) and the sample magnetization (\SI[separate-uncertainty]{1.0\pm0.2}{m\ampere}).}
\end{figure}

Finally, we demonstrate the effectiveness of our parabolic tips for scanning magnetometry by performing measurements of an out-of plane magnetized ferromagnet, specifically a \SI{1}{\nm} thick, \SI{0.73}{\micro\meter} wide stripe of CoFeB capped by a \SI{5}{\nm} layer of Ta (Fig.~\ref{fig:Linecut}).  The stripe is mounted in a home-built atomic-force and confocal microscope~\cite{appel_nanoscale_2015,appel_fabrication_2016}, with one of our parabolic diamond scanning probes.  We apply an external field of \SI{28}{G} along the NV axis to split the $m_s=\ket{\pm1}$ spin states.  We then perform linescans perpendicular to the stripe, maintaining a constant NV-sample separation via atomic-force feedback (Fig.~\ref{fig:Linecut}(a)), and record an ODMR spectrum every 20 nm along the scan.  We determine the stray field of the sample projected along the NV axis (Fig.~\ref{fig:Linecut}(b), blue dots) based on the Zeeman splitting of the ODMR resonances relative to the out-of-contact splitting.  We fit this stray field by a commonly used, analytic model for stray magnetic fields produced at such structures (Fig.~\ref{fig:Linecut}(b), red solid line)~\cite{tetienne_nature_2015}.  Based on this fit, we extract a sample magnetization of \SI[separate-uncertainty]{1.0\pm0.2}{m\ampere} and a separation of \SI[separate-uncertainty]{45\pm5}{\nm} between the NV and the CoFeB stripe. Since the CoFeB is capped by a \SI{5}{\nm} layer of Ta, the effective separation between NV and Ta surface is \SI[separate-uncertainty]{40\pm5}{\nm}. The truncated shape of our parabolic design minimizes this separation, leading to a spatial resolution comparable with the state of the art.  Furthermore, the increased PL signal from our devices, by a factor of 5 relative to similar measurements performed using cylindrical scanning probe devices\cite{maletinsky_robustsensor_2012,appel_fabrication_2016,zhou_scanning_2017}, directly leads to a corresponding reduction in measurement time.

\section{Conclusion}

In conclusion, we have fabricated parabolic diamond scanning probes containing single NV centers and demonstrated their use for nanoscale magnetic field imaging.  The parabolic design is ideal for sensing applications, as it yields a high rate of detected photons from a near-surface NV.  The devices could be further improved by incorporating an antireflection coating to the back surface of the cantilever, through the use of (111) oriented diamond to achieve optimal mode overlap with the NV optical transition dipoles~\cite{rohner_111-oriented_2019}, and better lateral NV placement via deterministic alignment to pre-selected NVs.  Our design is versatile and can be applied to many systems of interest, including scanning probe sensing of magnetic and electric fields or temperature, as well as NMR sensing of molecules or materials attached to the diamond surface~\cite{lovchinsky_nuclear_2016,lovchinsky_nqr_2017}.

\section{Acknowledgments}
We would like to thank M. Kasperczyk for fruitful discussions concerning the statistical analysis of our data. We further thank K. Garcia and R. Soucaille for fabrication of the CoFeB sample investigated. We gratefully acknowledge financial support through the NCCR QSIT (Grant No. 185902), the Swiss Nanoscience Institute, the EU Quantum Flagship project ASTERIQS (Grant No. 820394), and through the Swiss NSF Project Grants Nos.~188521 and 155845 and Spark Grant No.~190592.

N.H. conducted the majority of characterization and magnetometry measurements.  B.J.S. conceived and oversaw the project, developed the fabrication and characterization methods, and performed simulations and measurements.  B.J.S. and N.H. contributed equally to the fabrication of devices. D.R. and M.B. assisted with measurements. P.M. contributed to experimental methods.  All authors prepared the manuscript.  

%\bibliographystyle{apsrev4-1}
%\bibliography{ParabolicReflector}


\end{document}